\documentclass[bibyear]{aa}
\usepackage[varg]{txfonts}
\usepackage{graphicx}
\def\kms{\mbox{km~s$^{-1}$}}

\def\mic{\mbox{$\mu$m}}

\def\d{{\rm d}}
\def\e{{\rm e}}

\def\Ro{R_{\rm o}}
\def\Ri{R_{\rm i}}
\def\Yo{Y_{\rm o}}
\def\Yi{Y_{\rm i}}
\def\yo{y_{\rm o}}
\def\yi{y_{\rm i}}
\def\zo{z_{\rm M}}
\def\zi{z_{\rm m}}
\def\ri{r_{\rm i}}
\def\To{T_{\rm o}}
\def\Omo{\Omega_{\rm o}}
\def\snt{\sigma_{\rm NT}}
\def\so{\sigma_{\rm o}}
\def\sr{\eta}
\def\Mv{M_{\rm vir}}
\def\Mnt{M_{\rm NT}}

\def\Moth{M_{\rm o}^{\rm thin}}
\def\Morjth{M_{\rm o}^{\rm RJ-thin}}
\def\Mrj{M^{\rm RJ}}
\def\Mrjth{M^{\rm RJ-thin}}
\def\tho{\theta_{\rm o}}
\def\rhoo{\rho_{\rm o}}
\def\tauo{\tau_{\rm o}}
\def\mc{m_{\rm c}}
\def\RR{\mathcal{R}}
\def\dSnu{\delta S_\nu}
\def\Ns{N_{\rm S}}

\begin{document}
\title{
The mass of dusty clumps with temperature and density structure
}
\author{
        R.~Cesaroni
}
\institute{
 INAF, Osservatorio Astrofisico di Arcetri, Largo E. Fermi 5, I-50125 Firenze, Italy
	   \email{cesa@arcetri.astro.it}
}
\offprints{R. Cesaroni, \email{cesa@arcetri.astro.it}}
\date{Received date / Accepted date}

\abstract{
We consider a dusty clump in the two cases of spherical and cylindrical symmetry to
investigate the effect of temperature and density gradients on the observed flux density.
Conversely, we evaluate how the presence of such gradients affects the calculation of
the clump mass from the observed flux. We provide the reader with approximate expressions
relating flux density and mass in the optically thick and thin limits, in the Rayleigh-Jeans regime, and discuss the
reliability of these expressions by comparing them to the outcome of a numerical code.
Finally, we present an application of our calculations to three examples taken from the
literature, which shows how the correction introduced after taking into account
temperature and density gradients may affect our conclusions on the stability of the clumps.
}

\maketitle

\section{Introduction}
\label{sint}

Estimating the mass of molecular, dusty clumps is of great importance for
a number of reasons, such as the determination of the clump mass function,
the calculation of the virial parameter, the estimate of molecular abundances,
etc.. While various methods can be used for this purpose, the most common
takes advantage of the fact that the spectral energy distribution (SED) of
the continuum emission from a dusty, homogeneous, isothermal cloud can
be approximated as a modified black-body. In this case, the emission at
sufficiently long wavelengths is optically thin and the integrated flux
density can be easily expressed as a function of the mass and temperature
of the dust. For a given gas-to-dust mass ratio, this allows to derive
the total mass of the cloud from the flux density, if the dust temperature
and absorption coefficients are known.  In practice, the cloud mass is
evaluated as described in the pioneering study of Hildebrand (\cite{hild83})
and can be expressed as (see e.g. Eq.~(1) of Schuller et al.~\cite{schul09})
\begin{equation}
M = \frac{S_\nu d^2\,\RR}{\kappa(\nu) B_\nu(T)}   \label{eone}
\end{equation}
where $S_\nu$ is the flux density, $d$ the distance to the cloud, $B_\nu$
the Planck function, $T$ the dust temperature, $\kappa$ the dust
absorption coefficient per unit mass, and $\RR$ the gas-to-dust mass ratio.

While this simplified expression is perfectly adequate to most purposes,
real life is much more complicated. Observations are currently performed
at higher and higher frequencies, e.g from space with the Herschel Space
Observatory and from ground with the Atacama Large Millimeter and
submillimeter Array (ALMA) which is now operative up to 900~GHz. At such
bands dust optical depth cannot be neglected a priori and should be considered
when converting flux density into mass. Also, compact molecular cores
can be heated from outside (due to nearby luminous stars) or inside (due
to embedded forming stars), which generates temperature gradients that
in turn break the assumption of isothermal clump. Density gradients are
likely present too, owing to collapse during the star formation process
or other phenomena (e.g. expansion in molecular outflows).

Additional sources of uncertainty on the estimate of the mass are related
to the error on the flux measurement, the distance of the source
(often poorly known), and the value of the dust absorption coefficient,
which depends on the properties of the dust grains (see e.g. Ossenkopf \&
Henning~\cite{osshen}). The combination of all these errors may overcome
the error caused by the assumptions of low optical depth and constant
temperature. However, in some cases one is interested in quantities that
do not depend on distance (e.g. the mass-to-luminosity ratio)
or all targets are located basically at the same distance (as in studies
of the core mass function within the same molecular cloud),
which
makes the distance error irrelevant. In addition, for other quantities,
such as the virial parameter, it is important to determine whether the
value lies above a given threshold and is thus useful to improve on the
accuracy of the estimated parameter as much as possible. Neglecting the
opacity as well as the temperature and density gradients may lead to wrong
conclusions in these cases.

The goal of our study is to quantify the effects
of large dust opacity and temperature and density gradients on the clump
mass estimated with Eq.~(\ref{eone}). In particular, in Sect.~\ref{score}
we analyse the case of a spherically symmetric clump with temperature and
density varying as power laws of the radius, in Sect.~\ref{sdisk} we repeat
the same exercise for a cylindrically symmetric clump
and in Sect.~\ref{sappl} we apply the corrections estimated with
our method to data from the literature. Finally, the results are summarized
in Sect.~\ref{scon}.

\section{Flux density of spherical clump}
\label{score}

We want to calculate the integrated flux density emitted by the dust
in a spherically symmetric clump. In our model the gas and dust are
distributed between an inner radius $\Ri$ and an outer radius $\Ro$, and the
mass ratio between gas and dust, $\RR$, does not depend on the radius, $R$.
The dust temperature and density are expressed as
\begin{eqnarray}
T & = & \To \left(\frac{R}{\Ro}\right)^q    \label{etr} \\
\rho & = & \rhoo \left(\frac{R}{\Ro}\right)^p   \label{erh}
\end{eqnarray}
where $\To$ and $\rhoo$ are
the dust temperature and density
at the outer radius. By definition, the gas density is equal to $\rho \RR$.

\subsection{Approximate analytical expression}

As a first step, it is instructive to calculate the expression of the integrated
flux density in the optically thin and thick limits. In the latter, only the photons
emitted from the clump surface contribute to the observed flux, which is given by
\begin{equation}
 S_\nu = \frac{\pi B_\nu(\To) \, 4\pi\Ro^2}{4\pi d^2} = \Omo \, B_\nu(\To)   \label{ethick}
\end{equation}
with $\Omo=\pi\Ro^2/d^2$ solid angle subtended by the clump.
In practice such a thick limit can hardly be reached at (sub)millimeter wavelengths.
This can be seen by estimating the density needed to achieve a dust opacity of $1$
in a thin surface layer of thickness, e.g., $\Delta R=0.1\,\Ro$. It is easy to show that
the condition $\tau=\kappa \, \rhoo \, \Delta R=1$ in the template case
$p=0$ and $\ri=0$ can be re-written as
\begin{equation}
 \Sigma = \frac{4}{3} \frac{\RR}{\kappa} \frac{\Ro}{\Delta R}
\end{equation}
where $\Sigma=(4/3)\RR \rhoo \Ro$ is the mean surface density of the clump.
At 1~mm $\kappa\simeq1$~cm$^2$g$^{-1}$ (see Ossenkopf \& Henning~\cite{osshen})
and for $\RR=100$ one obtains $\Sigma\simeq10^3$~g~cm$^{-2}$, as opposed to
$\Sigma\la1$~g~cm$^{-2}$ of typical molecular clumps.

In the optically thin limit, instead, all photons
emitted by the grains freely escape from the clump and $S_\nu$ is
obtained from
\begin{equation}
S_\nu = \frac{1}{4\pi d^2} \int_{\Ri}^{\Ro} 4\pi j_\nu \, 4\pi R^2 {\d}R = \frac{4\pi}{d^2} \int_{\Ri}^{\Ro} \kappa \rho B_\nu R^2 \d R
\end{equation}
where $j_\nu$ is the dust emissivity and we made use of Kirchhoff's law $j_\nu/\kappa=B_\nu(T)$. If
$h\nu\ll kT$ (with $k$ Boltzmann constant and $h$ Planck constant),
this equation can be re-written using the Rayleigh-Jeans (hereafter RJ)
approximation:
\begin{eqnarray}
S_\nu & \simeq & \frac{4\pi}{d^2} \kappa \rhoo \frac{2k\nu^2}{c^2}\To
             \int_{\Ri}^{\Ro} R^2 \left(\frac{R}{\Ro}\right)^{q+p} \d R   \nonumber \\
      & = &  \frac{4\pi\Ro^3}{d^2} \kappa \rhoo \frac{2k\nu^2}{c^2}\To
             \int_{\ri}^1 r^{q+p+2} \d r    \label{esnu}
\end{eqnarray}
where we have defined $r=R/\Ro$ and $\ri=\Ri/\Ro$.

One can relate the expression of $S_\nu$ to the mass of the clump, $M$. The latter can be computed from
\begin{equation}
M = \int_{\Ri}^{\Ro} \rho \RR \, 4\pi R^2 \, \d R  =  4\pi \Ro^3 \, \rhoo \RR \int_{\ri}^1 r^{p+2} \, \d r.   \label{emass}
\end{equation}
From this expression and Eq.~(\ref{esnu}), one obtains
\begin{equation}
S_\nu = \frac{\kappa M}{\RR\,d^2} \frac{2k\nu^2}{c^2}\To \frac{\int_{\ri}^1 r^{q+p+2} \, \d r}{\int_{\ri}^1 r^{p+2} \, \d r} =
        m \frac{2k\nu^2}{c^2}\To\, F(\ri;q,a)   \label{ethin}
\end{equation}
where we have defined $m=\kappa M/(\RR\,d^2)$, $a=p+3$, and
$F=\int_{\ri}^1 r^{q+a-1} \, \d r \left/ \int_{\ri}^1 r^{a-1} \, \d r \right.$.
It is straightforward to demonstrate that function $F$ takes the following values:
\begin{equation}
F = \left\{
\begin{array}{lcl}
1 & \Leftrightarrow & q=0 \\
\frac{\ri^q-1}{\ln\ri^q} & \Leftrightarrow & q\ne0,~a=0 \\
\frac{\ln\ri^{-q}}{\ri^{-q}-1} & \Leftrightarrow & q\ne0,~a\ne0,~a=-q \\
\frac{a}{a+q} \frac{1-\ri^{a+q}}{1-\ri^a} & \Leftrightarrow & q\ne0,~a\ne0,~a\ne-q
\label{eff}
\end{array}
\right.
\end{equation}

Finally, from Eq.~(\ref{ethin}) one obtains
\begin{equation}
 m = \frac{S_\nu}{\frac{2k\nu^2}{c^2}\To} \frac{1}{F}  \label{ems}
\end{equation}
or, equivalently,
\begin{equation}
 M = \frac{S_\nu d^2 \, \RR}{\kappa(\nu) \frac{2k\nu^2}{c^2}\To} \frac{1}{F}  \label{emms}
\end{equation}
which is analogous to Eq.~(\ref{eone}) when the temperature and density gradients are taken into account.
We stress that these equations are valid only in the optically thin limit and under the RJ
approximation.

It is interesting to discuss the transition between optically thin and
optically thick regimes. The critical value of $m$ for which such
a transition occurs is obtained by equating the flux density from
Eq.~(\ref{ethick}), in the RJ limit, to that from Eq.~(\ref{ethin}):
\begin{equation}
 \mc = \frac{\Omo}{F}.   \label{emc}
\end{equation}
We note that the (approximate) relationship between $S_\nu$ and $m$ is
fully determined by Eqs.~(\ref{ethick}) and~(\ref{emc}), because once the
optically thick flux and the critical value of $m$ are fixed, also the
optically thin flux is univocally established.  This fact can be used to
study the dependence of $S_\nu$ on the various physical parameters.

\begin{figure}
\centering
\resizebox{8.5cm}{!}{\includegraphics[angle=0]{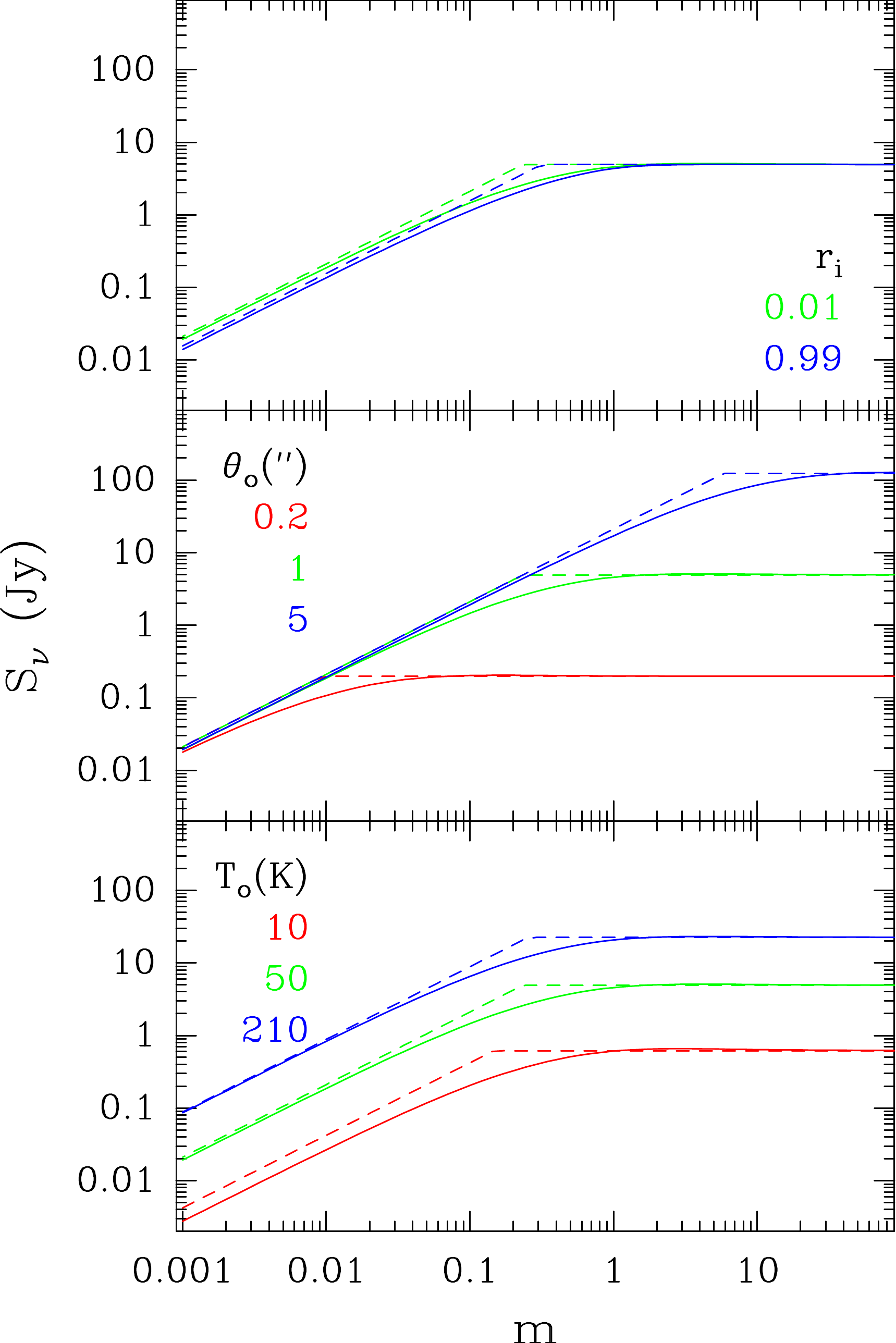}}
\caption{
Template flux densities from a spherical dusty clump as a function of
parameter $m$ (see text). The curves are obtained for illustrative
purposes from fiducial values of the input parameters, i.e. 
$\nu=220$~GHz, $\tho=\Ro/d=1\arcsec$, $\ri=0.01$, $\To=50$~K,
$q=-0.4$, $p=-1.5$. In each panel only one of these parameters
is changed as indicated in the panel itself. Dashed curves represent
the approximate analytical solutions given by Eq.~(\ref{ethin}),
while solid curves are obtained from the numerical model described
in Sect.~\ref{snume}.
}
\label{fsm}
\end{figure}

The behaviour of $S_\nu$ as a function of $m$ is illustrated by the dashed curves
in Fig.~\ref{fsm}. In all panels the green curve corresponds to the approximate expressions of
$S_\nu$ for a set of parameters arbitrarily chosen for illustrative
purposes. These are $\nu=220$~GHz, $\tho=\Ro/d=1\arcsec$, $\ri=0.01$, $\To=50$~K,
$q=-0.4$, $p=-1.5$ (i.e. $a=1.5$). The blue and red curves are obtained
by varying only one of these parameters, as detailed in each panel.

In particular, we observe that both the optically thick flux from
Eq.~(\ref{ethick}) and $\mc$ are proportional to $\Omo$, but only the former
depends on $\To$. This implies that for increasing $\To$ the flux density
increases, while the transition between the thin and thick regimes occurs
approximately\footnote{The slight shift of $\mc$ of the red curve in the
bottom panel of Fig.~\ref{fsm} is due to the RJ approximation being unsuited
for $\To=10$~K and $\nu=220$~GHz.}
at the same value of $m$. Instead, for increasing $\Omo$ both
the thick flux and $\mc$ increase by the same factor, while the optically
thin flux remains the same (because Eq.~(\ref{ethin}) does not depend on
$\Ro$). Finally, it can be shown that function $F$ is increasing with $\ri$
if $q>0$ and decreasing if $q<0$ (see Appendix~\ref{afri}),
which in turn implies that a variation of $\ri$ affects only $\mc$
and not the optically thick flux density.

The solid curves in the figure represent the flux density computed with
the numerical model described in the next section, which properly takes into
account the dust optical depth and does not assume the RJ approximation.

\subsection{Numerical solution}
\label{snume}

It is possible to obtain an exact semi-analytical expression
of $S_\nu$ as a function of the clump
mass only in the simple case $q=0$ and $p=0$.
The result is given by Eq.~(A.4) of Cesaroni et al.~(\cite{cesa19}),
which with our notation takes the form
\begin{eqnarray}
S_\nu & = & \Omo \, B_\nu(T) \nonumber \\
& &  \times \left[ 1+\frac{2}{\tauo} \left(\sqrt{1-\ri^2} \,
     \e^{-\tauo\sqrt{1-\ri^2}} +
     \frac{\e^{-\tauo\sqrt{1-\ri^2}}-1}{\tauo}\right) \right.   \label{eexa} \\
& &  \left. ~~~~- \int_0^{\ri^2} \e^{-\tauo\left(\sqrt{1-t}-\sqrt{\ri^2-t}\right)} {\rm d}t \right] \nonumber
\end{eqnarray}
where $\tauo=2\Ro\,\kappa \rhoo=3m/(2\Omo)$.

More in general, the clump flux density can be estimated numerically as
the integral of the brightness $I_\nu$ over the source solid angle, namely
\begin{equation}
 S_\nu = \int_{\Omo} I_\nu {\rm d}\Omega
       = \frac{1}{d^2} \int_0^{\Ro} I_\nu(x)\,2\pi x {\rm d}x
       = 2 \Omo \int_0^1 I_\nu(\xi) \, \xi \, \d\xi
 \label{enum}
\end{equation}
where $x$ is the projected radius on the plane of the sky and we assume $\xi=x/\Ro$.

\begin{figure}
\centering
\resizebox{8.5cm}{!}{\includegraphics[angle=0]{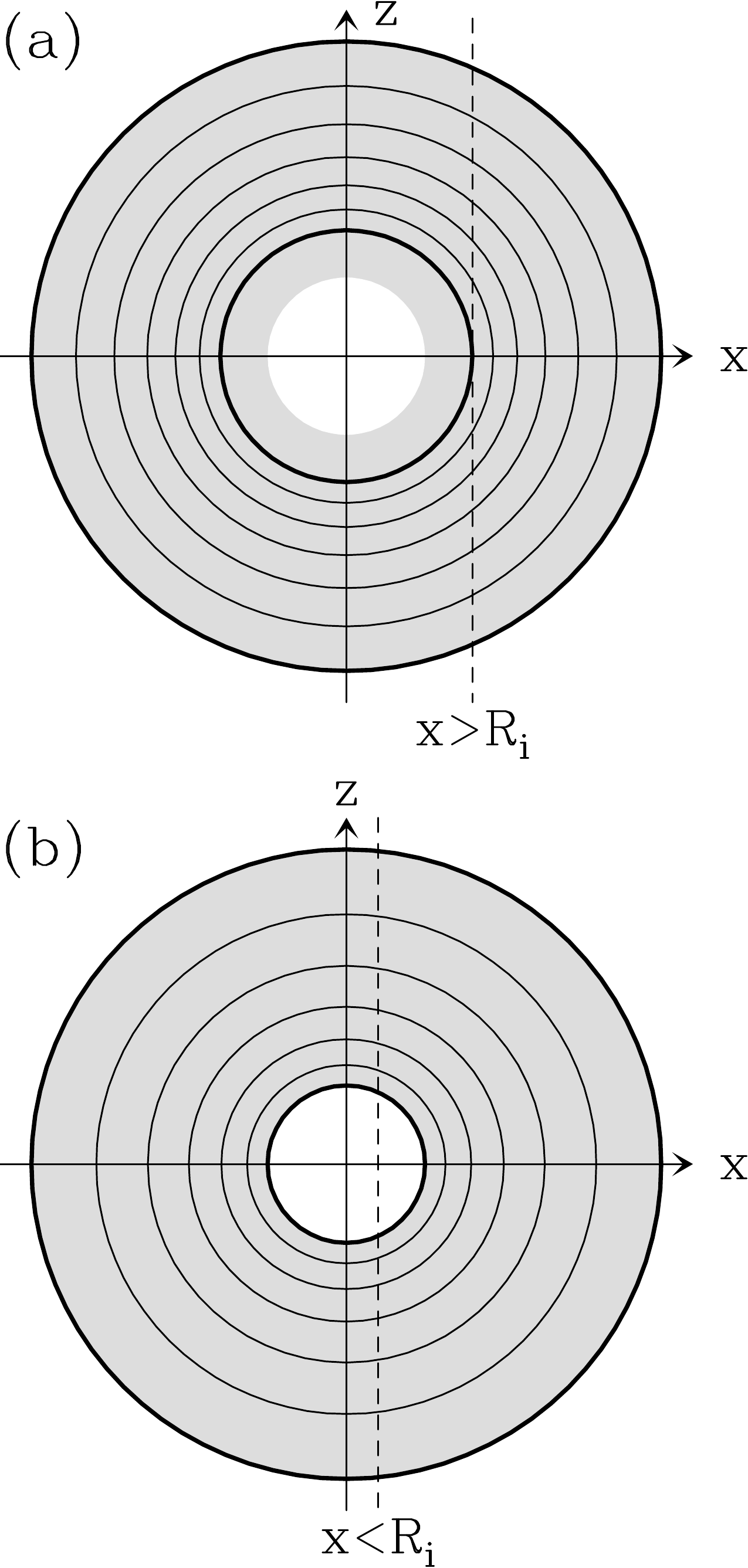}}
\caption{
Sketch of two lines of sight (dashed lines) through a spherically
symmetric clump (grey area) with impact parameter $x$, crossing (bottom
panel) and not crossing (top panel) the central cavity of radius $\Ri$
(white central area). The thick circles denote the minimum and maximum
radius of the region contributing to the brightness along the given l.o.s.,
while the thin circles are template annuli defined by Eq.~(\ref{erj}),
to be used for the numerical integration of the radiative transfer equation.
}
\label{fske}
\end{figure}

It is convenient to split the calculation of the brightness
along an arbitrary l.o.s. through the clump into two parts, for positive and negative values of $z$, as follows:
\begin{eqnarray}
 I_\nu^0 & = & I_\nu^{\rm BG} \, \e^{-\int_{\zi}^{\zo} \kappa\rho\,\d z} \, + \, \int_{\zi}^{\zo} B_\nu \, \e^{-\int_{\zi}^z \kappa\rho\,\d z'} \kappa\rho\,\d z   \label{einup} \\
 I_\nu & = & I_\nu^0 \, \e^{-\int_{-\zo}^{-\zi} \kappa\rho\,\d z} \, + \, \int_{-\zo}^{-\zi} B_\nu \, \e^{-\int_{-\zo}^z \kappa\rho\,\d z'} \kappa\rho\,\d z   \label{einun}
\end{eqnarray}
where $z$ is the Cartesian coordinate along the line of sight (l.o.s.),
$I_\nu^{\rm BG}$ is the background brightness, the observer is located at
$z=-\infty$, and we define
\begin{eqnarray}
 \zo & = & \sqrt{\Ro^2-x^2} \\
 \zi & = & 
 \left\{
 \begin{array}{lcl}
  \sqrt{\ri^2-x^2} & \Leftrightarrow & 0\le x<\ri \\
  0 & \Leftrightarrow & \ri\le x\le1
 \end{array}
 \right.
\end{eqnarray}
(see Figs.~\ref{fske}a and~\ref{fske}b for a sketch of two representative l.o.s.).

In the following we focus on the solution of Eq.~(\ref{einup}). The emergent
brightness at $z=-\zo$ given by Eq.~(\ref{einun}) can be calculated with
the same approach described below, once $I_\nu^0$ has been computed.

In order to obtain an approximate analytical solution of Eq.~(\ref{einup}),
we divide the part of the clump which contributes to the radiation along
the given l.o.s. into a suitable number of shells, $\Ns$, and assume that in
each shell the relevant physical parameters (density and temperature)
are constant. Figure~\ref{fske}a shows a sketch of the shells for a
generic l.o.s. with $x>\Ri$, where only the dust between $R=x$ and
$R=\Ro$ contributes to the brightness, while Fig.~\ref{fske}b refers to
the l.o.s. with $0\le x<\Ri$, where the portion contributing to $I_\nu$
is the whole shell between $R=\Ri$ and $R=\Ro$.

Under the previous approximation, Eq.~(\ref{einup}) takes the form
\begin{eqnarray}
 I_\nu^0 & = & I_\nu^{\rm BG} \, \e^{-\sum_{j=1}^{\Ns} \int_{z_{j-1}}^{z_j} \kappa\rho \, \d z} \nonumber \\
     & &   + \sum_{j=1}^{\Ns} \int_{z_{j-1}}^{z_j} B_\nu(T) \,
	 \e^{-\int_{\zi}^{z_{j-1}} \kappa\rho \, \d z' - \int_{z_{j-1}}^z \kappa\rho \, \d z'} \kappa\rho \, \d z  \nonumber \\
     & \simeq & I_\nu^{\rm BG} \, \e^{-\sum_{j=1}^{\Ns} \kappa\rho_j(z_j-z_{j-1})} \nonumber \\
     & &   + \sum_{j=1}^{\Ns} B_\nu(T_j) \,
	 \e^{-\sum_{l=1}^{j-1} \kappa\rho_l (z_l-z_{l-1})} \kappa\rho_j
         \int_{z_{j-1}}^{z_j} \e^{-\kappa\rho_j(z-z_{j-1})} \, \d z \nonumber \\
    & = & I_\nu^{\rm BG} \e^{-\sum_{j=1}^{\Ns} \tau_j}
           + \sum_{j=1}^{\Ns} B_\nu(T_j) \left(1-\e^{\tau_j}\right) \, \e^{-\sum_{l=1}^{j-1} \tau_l}
 \label{esol}
\end{eqnarray}
where we define $z_0=\zi$, $\sum_{l=1}^0\tau_l=0$,
$\tau_j=\kappa\rho_j(z_j-z_{j-1})$, $T_j=T(R_j)$, and $\rho_j=\rho(R_j)$,
with $R_j$ outer radius of shell $j$.
The opacity of shell $j$ can be written as
\begin{eqnarray}
 \tau_j & = & \kappa\rhoo\Ro \,r_j^p \, \left(\sqrt{r_j^2-\xi^2}-\sqrt{r_{j-1}^2-\xi^2}\right)   \nonumber \\
        & = & \frac{m}{4\Omo} \frac{r_j^{\,a-3}}{\int_{\ri}^1 r^{\,a-1} \d r} \left(\sqrt{r_j^2-\xi^2}-\sqrt{r_{j-1}^2-\xi^2}\right)
\end{eqnarray}
where we used Eq.~(\ref{emass}).

Equation~(\ref{esol}) can be easily implemented in a computer code as it is
equivalent to iteratively solving the radiative transfer equation for
each shell, using as input the output brightness of the previous shell
crossed by the l.o.s..

The major problem with this approach is that a priori both the density
and/or temperature laws may be very steep close to the clump center, if $q$ and/or
$p$ are negative. Therefore, the thickness of the shells cannot be constant
and must be adapted to the local value of the density and temperature
gradients. We propose a simple way to get around this problem.

In practice, what matters for our purposes is to estimate the flux density
to a desired level of accuracy, $\dSnu$. This means that we should divide
the clump into a number of shells, $\Ns$, such that each of them does not
contribute more than $\dSnu$ to the total flux density. For a given
l.o.s. with impact parameter $x$, the shells to be considered in
Eq.~(\ref{esol}) are those with $R\ge x$, if $x>\Ri$, and $R\ge\Ri$, if $x\le\Ri$
(see Fig.~\ref{fske}). Thus the total flux density of interest for the
integration along the given l.o.s. is that emitted between
$r=r_0={\rm max}\{\xi,\ri\}$ and $r=1$.
This implies that a suitable value of $\Ns$ is given by
\begin{equation}
 \Ns = \left[\frac{S_\nu(r_0;1)}{\dSnu}\right]+1.   \label{ensh}
\end{equation}
Here the square brackets indicate the integer part of the argument and
1 is added to prevent the case $\Ns=0$.  Moreover, we use the notation
$S_\nu(r_1;r_2)$ to indicate the flux density emitted between two generic
radii $R_1<R_2$, which implies that $S_\nu(\ri;1)$ is the total flux
density emitted by the clump.

The expression for the radius of a generic shell, $j$, is derived by imposing that
each shell equally contributes with a fraction $1/\Ns$ to the total flux density $S_\nu(r_0;1)$, namely
\begin{equation}
 S_\nu(r_{j-1};r_j) = \frac{S_\nu(r_0;1)}{\Ns}   \label{esjj}
\end{equation}
for any $j=1,\dots,\Ns$, under the assumption that $r_j>r_{j-1}$.

An approximate expression of $S_\nu(r_1;r_2)$, with $r_1<r_2$, can be calculated
in the optically thin and RJ limits from Eq.~(\ref{esnu}):
\begin{equation}
 S_\nu(r_1;r_2) \propto \int_{r_1}^{r_2} r^{\,a+q-1} \d r =
 \left\{
 \begin{array}{lcl}
  \frac{r_2^{a+q}-r_1^{a+q}}{a+q} & \Leftrightarrow & a+q\ne0 \\
  \ln\left(\frac{r_2}{r_1}\right) & \Leftrightarrow & a+q=0
 \label{esrr}
 \end{array}
 \right.
\end{equation}
Substituting this expression in Eq.~(\ref{esjj}), one obtains
\begin{equation}
 r_j^{a+q} = r_{j-1}^{a+q} + \frac{1-r_0^{a+q}}{\Ns}
\end{equation}
for $a+q\ne0$, and
\begin{equation}
 \ln r_j = \ln r_{j-1} - \frac{\ln r_0}{\Ns}
\end{equation}
for $a+q=0$.
Since these expressions hold for any $j$, after some algebra one can finally write
\begin{equation}
 r_j =
 \left\{
 \begin{array}{lcl}
  \left(r_0^{a+q} + j\frac{1-r_0^{a+q}}{\Ns}\right)^{\frac{1}{a+q}}& \Leftrightarrow & a+q\ne0 \\
  r_0^{1-\frac{j}{\Ns}} & \Leftrightarrow & a+q=0.
 \label{erj}
 \end{array}
 \right.
\end{equation}

Using Eq.~(\ref{esrr}) and setting $\delta S_\nu=\varepsilon S_\nu(\ri;1)$,
one can also conveniently re-write Eq.~(\ref{ensh}) as
\begin{equation}
 \Ns = \left[\frac{S_\nu(r_0;1)}{\varepsilon S_\nu(\ri;1)}\right]+1 = \left[\frac{1}{\varepsilon} \frac{1-r_0^{a+q}}{1-\ri^{a+q}}\right] + 1   \label{enshn}
\end{equation}
where $\varepsilon$ is the fraction of the total flux density emitted by
the clump that we want to be contributed by each shell.

The solid curves in Fig.~\ref{fsm} are the numerical solutions obtained
for the same set of parameters as the dashed curves with the same colour.
For the sake of simplicity, in our calculations we have assumed $I_\nu^{\rm
BG}=0$.  While, as expected, the numerical solution tends to converge
to the corresponding approximate analytical solution for large and small
values of $m$, the two may differ significantly for intermediate values
of $m$. Moreover, some difference is also seen at small values of $m$
due to the RJ approximation. In Sect.~\ref{sdiff} we discuss all these
features in more detail.

\subsection{Limits of the approximate analytical solutions}
\label{sdiff}

\begin{figure}
\centering
\resizebox{8.5cm}{!}{\includegraphics[angle=0]{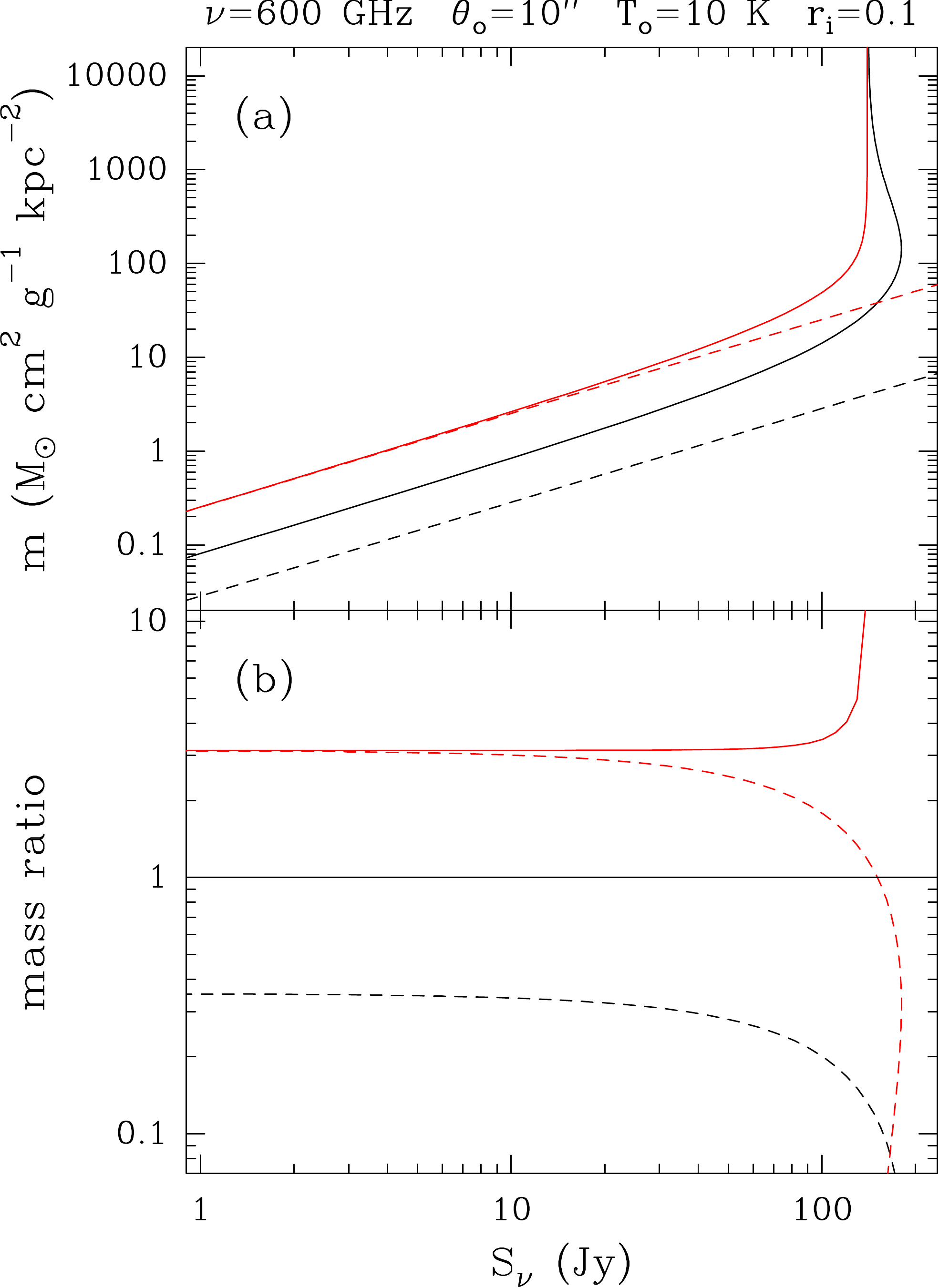}}
\caption{
{\it Panel a}: Plot of $m$ versus the total flux density of the clump.
The red curves correspond to the case $q=0$ and $p=0$, while the black
curves are for models allowing for temperature and density gradients.
The dashed black curve has been obtained under the optically thin and RJ approximations
from Eq.~(\ref{ems}), whereas the dashed red curve is computed in the
optically thin limit from Eq.~(\ref{eone}).
{\it Panel b}: Mass ratios between all the curves in the top panel and the black solid
curve.
}
\label{fcomp}
\end{figure}

The main goal of our study is to establish how much the conversion
from flux to mass can be affected by the usual assumption of constant
dust density and temperature. Therefore, it is convenient to consider the
inverse relationship with respect to those in Fig.~\ref{fsm} and plot the
core mass as a function of the flux density. With this in mind, in Fig.~\ref{fcomp}a
we show a plot of $m$, our proxy for the clump mass, versus $S_\nu$. For illustrative
purposes, we have considered an extreme case with $\nu=600$~GHz, $\tho=10\arcsec$,
$\To=10$~K, $q=-0.5$, $p=-2$, and $\ri=0.1$, which emphasizes the drawbacks of
using an approximate solution, as we show later. This set of parameters
could represent a typical clump observed e.g. in the Hi-GAL survey at 500~\mic.

For the sake of comparison, in the same figure beside the numerical solution
(black solid curve) we plot also the approximate analytical solution in
the optically thin and RJ limits (black dashed curve) from Eq.~(\ref{ethin}),
and the relationships (red curves) obtained under the commonly used assumption
of constant density and temperature (equal to $\rhoo$ and $\To$,
respectively). In particular, the red solid curve corresponds to the
solution from Eq.~(\ref{eexa}) while the red dashed curve is computed in
the optically thin limit from Eq.~(\ref{eone}).

To emphasize the comparison between the various curves, in
Fig.~\ref{fcomp}b we plot the ratio between the masses derived under
the different approximations and that computed numerically. Clearly,
at low fluxes the optically thin approximation is valid, as demonstrated
by the excellent match between the solid and dashed red curves. However,
for the same fluxes one sees a significant difference between the solid
and dashed black curves, due to the RJ approximation. At high
fluxes the deviation with respect to the numerical solution is very
prominent until the emission saturates due to the large opacity and
a mass estimate cannot be obtained because of degeneracy of the
solution.

We remark that the above example is proposed only for illustrative purposes.
More in general, one must keep in mind that the deviation from the
correct solution is sensitive to the input parameters of the model. This
is especially true for the observing frequency and dust temperature, on
which the goodness of the RJ approximation depends, and the steepness of
the temperature and density gradients. The effect of such gradients can be
seen by taking the ratio in the optically thin and RJ limits between the
mass from Eq.~(\ref{emms}) and that from Eq.~(\ref{eone}). It is
straightforward to demonstrate that such a ratio is equal to $1/F$, which
depends only on $\ri$, $q$, and $p$ or, equivalently, $a$. This result relies
upon the assumption that the temperature used in Eq.~(\ref{eone}) is $\To$.
In fact, most studies derive the clump temperature from a modified
black-body fit to the SED of the source, which usually peaks in the far-IR,
where the emission is optically thick and traces the outer layers of the
clump. Therefore, the temperature thus derived is very close to $\To$.

\begin{figure}
\centering
\resizebox{8.5cm}{!}{\includegraphics[angle=0]{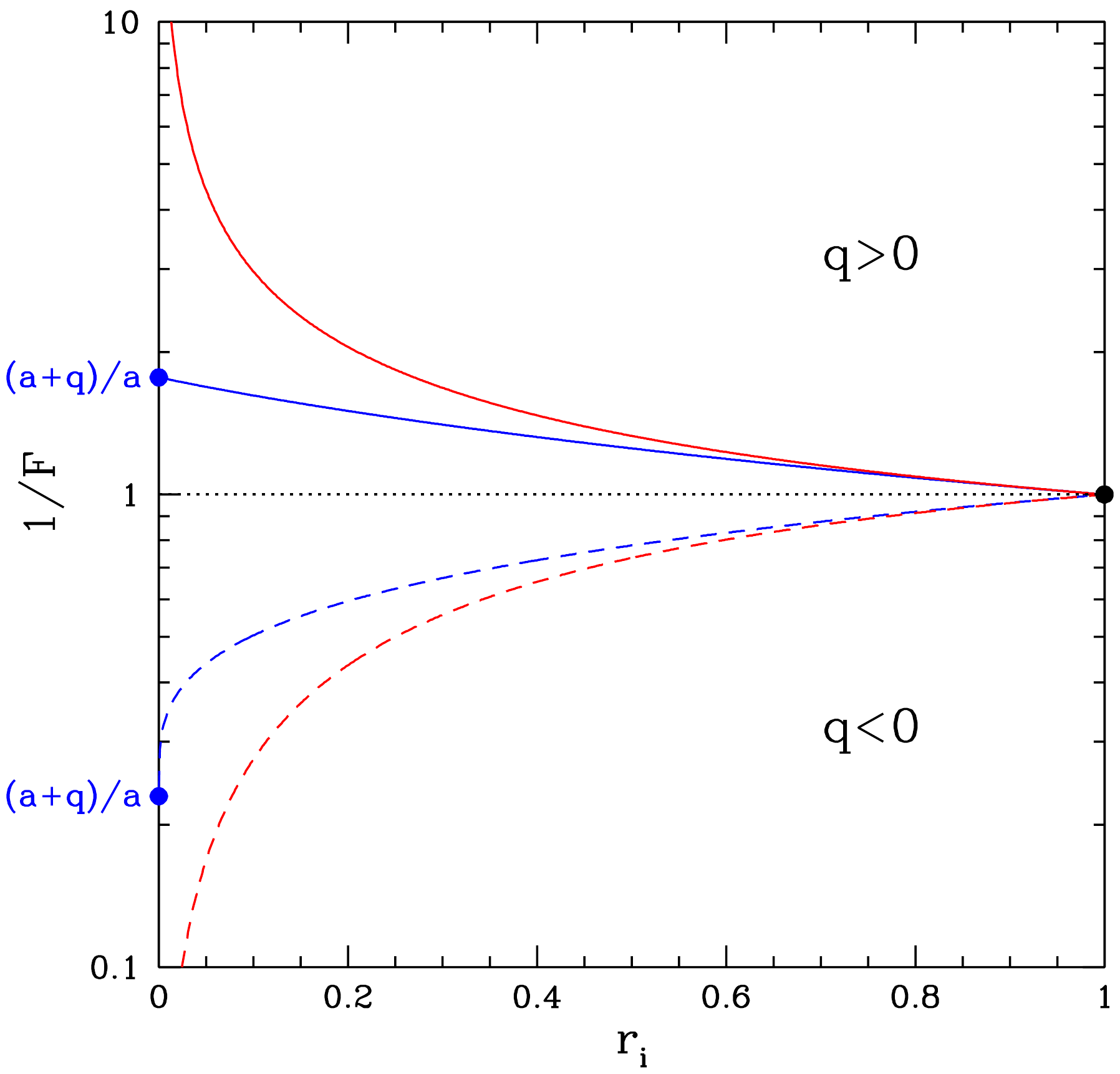}}
\caption{
Plot of $1/F$ as a function of $\ri$. Besides the trivial case $q=0$
(dotted line), four representative cases have been considered. Dashed
and solid lines correspond, respectively, to $q<0$ and $q>0$, while
blue indicates curves with $a+q>0$ and $a>0$ and red all the other cases.
The blue dots mark the value ($(a+q)/a$) of the corresponding curve for
$\ri\rightarrow0^+$ and the black dot indicates the limit (1) of all curves
for $\ri\rightarrow1^-$.
}
\label{ffinv}
\end{figure}

Figure~\ref{ffinv} shows the typical behaviour of $1/F$ as a function of
$\ri$ (see also Appendix~\ref{afri}), for $q=0$ (dotted line), $q\ne0$,
$a\ne0$, $a+q\ne0$ (blue curves), and in all the other cases (red
curves). One sees that a priori the presence of temperature and density
gradients may lead to largely underestimate (if $q>0$) or overestimate (if
$q<0$) the mass of the clump, for sufficiently small values of $\ri$. Whether
this occurs in practice and to what extent is discussed by means of a few
examples in Sect.~\ref{sappl}.

\section{Flux density of cylindrical clump}
\label{sdisk}

\begin{figure}
\centering
\resizebox{8.5cm}{!}{\includegraphics[angle=0]{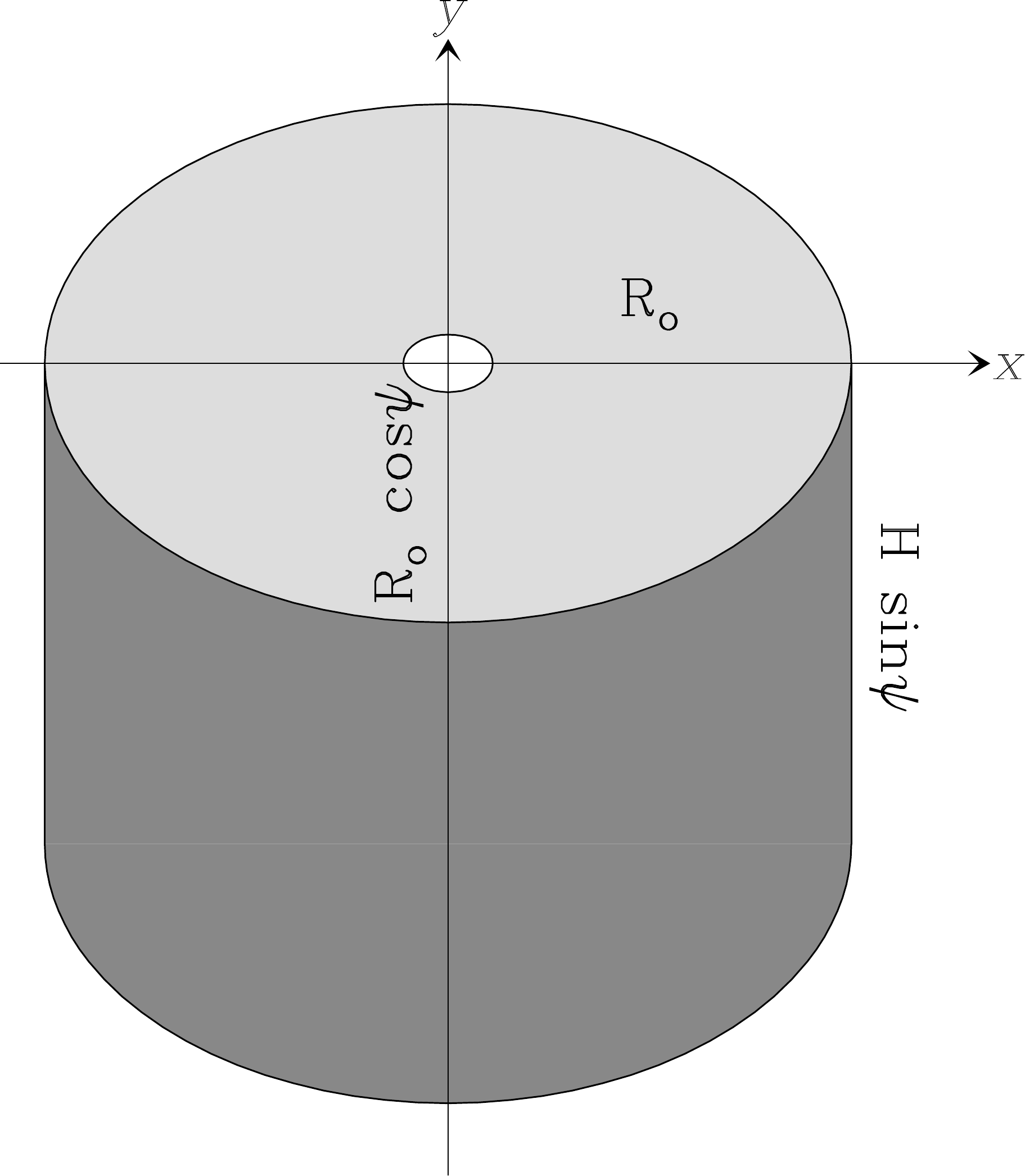}}
\caption{
Sketch of a cylindrical clump seen with an inclination angle $\psi$
between the symmetry axis and the l.o.s.. The figure represents the
projection of the cylinder on the plane of the sky, where the Cartesian
system $x,y$ lies. The radius and height of the cylinder are, respectively, $\Ro$ and $H$. 
}
\label{fdisk}
\end{figure}

Now, we compute the total flux density emerging from a cylindrically
symmetric clump with height $H$, inner radius $\Ri$, and outer radius $\Ro$.
This model might be more appropriate, e.g., for (part of) those filamentary structures
observed all over the Galaxy.
Figure~\ref{fdisk} shows the projection of the clump over the plane of the
sky for a generic inclination angle, $\psi$, between the l.o.s. and the
symmetry axis ($\psi=0$ corresponds to face on). Temperature and density depend only on $R$
through Eqs.~(\ref{etr}) and~(\ref{erh}).

\subsection{Approximate analytical expression}

As already done in Sect.~\ref{score}, it is instructive as a first step
to consider the solution in the optically thin and thick limits.

\subsubsection{Optically thick case}

If the
opacity is large, the flux is obtained by integrating the surface
brightness over the solid angle subtended by the source. This is the sum
of the integral over the light-grey and the dark-grey areas in Fig.~\ref{fdisk}.
The latter has constant brightness equal to $B_\nu(\To)$ and surface comprised
between two half ellipses described by the expressions
\begin{eqnarray}
 y_1 & = & \cos\psi \sqrt{\Ro^2-x^2} - H \, \sin\psi \\
 y_2 & = & \cos\psi \sqrt{\Ro^2-x^2}
\end{eqnarray}
where $x$ and $y$ are Cartesian coordinates lying in the plane of the
sky and oriented as shown in Fig.~\ref{fdisk}.
The flux density of such a surface is hence given by
\begin{equation}
 S_\nu^{\rm A} = \frac{B_\nu(\To)}{d^2} \int_{-\Ro}^{\Ro} \d x \int_{y_1(x)}^{y_2(x)} \d y 
 = \frac{2\Ro H}{d^2}B_\nu(\To) \sin\psi.
\end{equation}

The brightness over the light-grey ellipse in Fig.~\ref{fdisk} varies
with $R$ and the corresponding flux density is computed as follows:
\begin{eqnarray}
 S_\nu^{\rm B} & = & \frac{4}{d^2} \left[ \int_0^{\Ri} \d x \int_{\yi(x)}^{\yo(x)} B_\nu \, \d y +
                                          \int_{\Ri}^{\Ro} \d x \int_0^{\yo(x)} B_\nu \, \d y \right] \nonumber \\
 & = & \frac{4\cos\psi}{d^2} \left[ \int_0^{\Ri} \d X \int_{\Yi(X)}^{\Yo(X)} B_\nu(T(R)) \, \d Y \right. \nonumber \\
 &   & \left. +~\int_{\Ri}^{\Ro} \d X \int_0^{\Yo(X)} B_\nu(T(R)) \, \d Y \right]
 \label{esb}
\end{eqnarray}
where $\Yi=\sqrt{\Ri^2-x^2}$, $\Yo=\sqrt{\Ro^2-x^2}$, $\yi=\Yi\cos\psi$,
and $\yo=\Yo\cos\psi$, with $X,Y$ Cartesian coordinates perpendicular to the
cylinder axis, related to the $x,y$ system through the expressions $x=X$,
$y=Y\,\cos\psi$. In practice, Eq.~(\ref{esb}) is the integral of $B_\nu$
over the face of the cylinder, multiplied by $\cos\psi$. This integral is more conveniently expressed in
polar coordinates as
\begin{eqnarray}
 S_\nu^{\rm B} & = & \frac{4\cos\psi}{d^2} \int_{0}^{\frac{\pi}{2}} \d\phi \int_{\Ri}^{\Ro} B_\nu(T(R))\, R\, \d R \nonumber \\
 & = & 2\cos\psi\frac{\pi\Ro^2}{d^2} \int_{\ri}^1 B_\nu(T(r))\, r\, \d r
\end{eqnarray}

The total flux density is hence given by the sum $S_\nu^{\rm A}+S_\nu^{\rm B}$, namely
\begin{equation}
 S_\nu = \Omo^{\rm e} \sin\psi\, B_\nu(\To) + 2\,\Omo \cos\psi \int_{\ri}^1 B_\nu(T(r)) \, r \, \d r
\end{equation}
where $\Omo^{\rm e}=2\Ro\,H/d^2$ is the solid angle subtended by the clump seen edge on.
In the RJ approximation one obtains
\begin{equation}
 S_\nu \simeq \frac{2k\nu^2}{c^2}\To \left( \Omo^{\rm e} \sin\psi + 2\,\Omo \cos\psi \int_{\ri}^1 r^{q+1} \d r \right)
\end{equation}
with
\begin{equation}
 \int_{\ri}^1 r^{q+1} \d r =
 \left\{
 \begin{array}{lcl}
  \frac{1-\ri^{q+2}}{q+2} & \Leftrightarrow & q\ne-2 \\
  -\ln\ri & \Leftrightarrow & q=-2.
 \end{array}
 \right.
\end{equation}

\subsubsection{Optically thin case}

In the optically thin limit, the flux density does not depend on the
inclination angle because by definition the observer sees all the particles
of the clump that contribute to the photon budget, independently of the
shape and orientation of the clump. Therefore, the source luminosity is
computed by integrating the emissivity over the clump volume:
\begin{eqnarray}
 S_\nu & = & \frac{H}{4\pi d^2} \int_{\Ri}^{\Ro} 4\pi \kappa\,\rho(R)\,B_\nu(T(R))\, 2\pi R \,\d R \\
       & \simeq & 2 \frac{\pi\Ro^2H}{d^2} \rhoo\kappa \frac{2k\nu^2}{c^2}\To \int_{\ri}^1 r^{p+1} \d r.
\end{eqnarray}
where we have adopted the RJ approximation.
Since the mass of the clump is equal to
\begin{equation}
M = \int_{\Ri}^{\Ro} \rho(R) \RR\, 2\pi R \,H \,\d R  =  2\pi H\Ro^2 \, \rhoo \RR \int_{\ri}^1 r^{p+1} \d r
\end{equation}
one finally obtains
\begin{equation}
S_\nu = \frac{\kappa M}{\RR\,d^2} \frac{2k\nu^2}{c^2}\To \frac{\int_{\ri}^1 r^{\,q+p+1} \d r}{\int_{\ri}^1 r^{p+1} \d r} =
        m \frac{2k\nu^2}{c^2}\To\, F(\ri;q,a).
\end{equation}
This expression is formally identical to Eq.~(\ref{ethin}), with the only difference that this time
we have defined $a=p+2$.

\subsection{Numerical solution}

Now, we consider the general case with moderate opacity, which allows only
a numerical solution.  The calculation of the flux density for an arbitrary
inclination angle is quite complicated and goes beyond the scope of
the present study. Here, we consider only the two extreme inclinations:
face-on and edge-on.

\subsubsection{Edge-on cylindrical clump}

The calculation of $S_\nu$ is formally identical to that developed in Sect.~\ref{snume},
with the only difference that Eq.~(\ref{enum}) must be replaced with
\begin{equation}
 S_\nu = \int_{\Omo^{\rm e}} I_\nu \d\Omega
       = \frac{2}{d^2} \int_0^{\Ro} I_\nu(x)\, H \, \d x
       = \Omo^{\rm e} \int_0^1 I_\nu(\xi) \, \xi \, \d\xi.
\end{equation}

The brightness $I_\nu$ can be obtained by integrating along the l.o.s.
exactly as described in Sect.~\ref{snume}, provided $a=p+3$ is replaced
with $a=p+2$.

\subsubsection{Face-on cylindrical clump}

If the l.o.s. is parallel to the axis of the cylindrical clump, the flux density
is computed from Eq.~(\ref{enum}). The expression of the brightness, $I_\nu$, is easily
obtained because for a given $x$ the density and temperature are constant along the l.o.s.,
hence
\begin{equation}
I_\nu(\xi) = I_\nu^{\rm BG} \, \e^{-H\kappa\rhoo\,\xi^p} + B_\nu(\To\,\xi^q)\left(1-\e^{-H\kappa\rhoo\,\xi^p}\right)
\end{equation}
where we remind the reader that we have defined $\xi=x/\Ro$.

\section{Application to practical cases}
\label{sappl}

As a test bed for the clump model previously described, we consider three
examples taken from the literature. In two of these, the mass was estimated
under the usual hypothesis of constant temperature and density and assuming
optically thin emission.

\subsection{The hot molecular core G31.41+0.31}

\begin{figure}
\centering
\resizebox{8.5cm}{!}{\includegraphics[angle=0]{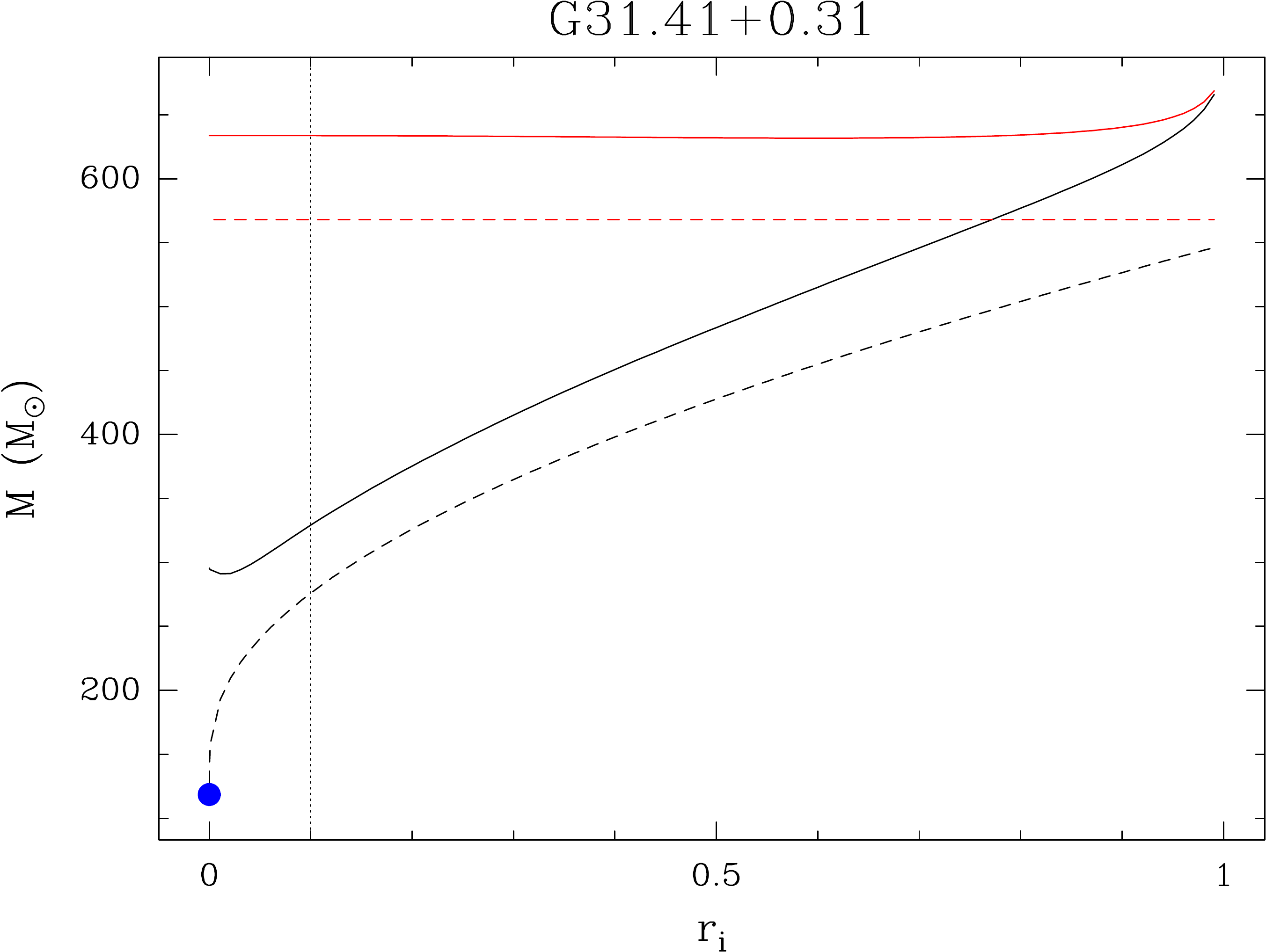}}
\caption{
Mass of the HMC G31.41+0.31 as a function of $\ri$. The input parameters are 
$d=7.9$~kpc, $\Ro=1\farcs076$, $q=-0.77$, $p=-2$, $\kappa(217{\rm GHz})=0.8$, $\RR=100$,
$S_{\rm 217\,GHz}=3.1$~Jy.
The black lines are for $q=-0.77$ and $p=-2$, whereas the red lines correspond to
$q=0$ and $p=0$.
Both dashed lines are obtained in the optically thin limit, while the
black dashed line assumes also the RJ approximation.
The blue dot corresponds to the value of the mass computed by BEL18.
The dotted vertical line marks a plausible upper limit for $\ri$ (see text).
}
\label{fgtu}
\end{figure}

As a first example, we consider the hot molecular core (HMC) G31.41+0.31, for which
Beltr\'an et al. (\cite{bel18}; hereafter BEL18) derived a mass estimate from the 1.4~mm
continuum emission imaged with ALMA. This case is especially suitable
for our purposes because these authors obtained also an estimate of the
temperature and density profiles as a function of the HMC radius.
We adopt the same parameters used in their calculation, namely $d=7.9$~kpc,
$\Ro=1\farcs076$, $q=-0.77$, $p=-2$, $\kappa(217{\rm GHz})=0.8$, and $\RR=100$.
The total flux density of the core at $\nu=217$~GHz is $S_\nu=3.1$~Jy.
The only unknown parameter is $\ri$, which BEL18
implicitly assumed equal to 0. In Fig.~\ref{fgtu} we plot the
values of the mass estimated in different ways, as a function of $\ri$.

The mass obtained from our numerical solution
(i.e. without any approximation)
is represented by the
black solid curve, while that derived under the optically thin and RJ
approximations is shown as a dashed black curve. For the sake of comparison,
we also mark with a blue dot the mass computed from Eq.~(6) of
BEL18. The resulting expression differs from our Eq.~(\ref{emms})
by only a factor $(2/\sqrt{\pi})[\Gamma(-(p+q)/2)/\Gamma(-(p+q+1)/2)]\simeq
0.927$. The latter is due to the fact that BEL18 calculated the brightness
by integrating along the line of sight from $-\infty$ to $+\infty$, whereas
we limit our integration to the sphere of radius $\Ro$.
Finally, we report in the same figure also the mass estimated
from Eq.~(\ref{eexa}) (i.e. without the RJ approximation and assuming
constant temperature and density)
both with (red dashed curve) and without (red solid curve) the optically
thin assumption.

The largest difference between the various curves occurs for $\ri=0$,
not surprisingly because at small radii the effect of the temperature
gradient is enhanced. Vice versa, for $\ri$ close to 1, the temperature
variation across the core is minimum and all curves converge towards the
$q=0$ solution corresponding to the red curves. In particular, the BEL18 solution
for $\ri=0$ is smaller than our numerical solution by a factor $\sim$2, whereas
the constant-temperature solutions predict a mass in excess by at least
a factor $\sim$2. It is also worth noting that the emission is partially thick
in this HMC, as proved by the gap between the solid curves and the corresponding
dashed curves.

The assumption $\ri=0$ is obviously unrealistic, as the temperature and
density laws must break down at some point close to the HMC center. A
plausible hypothesis is that $\Ri$ is comparable to half the separation
(0\farcs1) between the two free-free sources detected by Cesaroni et
al.~(\cite{cesa10}) close to the core center, which implies $\ri\simeq0.1$
(see dotted line in Fig.~\ref{fgtu}). For this value the discrepancy among
the different estimates of the mass is less prominent, but may still amount
to 70\%, which might not be negligible when comparing the core mass to
other parameters such as the virial mass or the magnetic critical mass.

\subsection{Stability of massive star-forming clumps}
\label{sfont}

Another convenient test-case for our model is represented by the sample of
massive clumps observed by Fontani et al.~(\cite{fonta02}; hereafter FON02).
In fact, also in this case as for BEL18 a direct estimate of the temperature
and density gradients was obtained by the authors, who find $q=-0.54$
and $p=-2.6$. A puzzling result of their study is that the
ratio between the clump masses and the corresponding virial masses is
$>$1 (see their Fig.~6), which hints at some additional support to stabilize the
clumps, such as e.g. magnetic fields. However, the mass estimates made by FON02
were derived without taking into account the temperature and density gradients
inside the clumps. Here, we want to reconsider the problem by applying the
appropriate corrections for these gradients.

At the time of FON02 no homogeneous data set was available for the continuum
emission of the clumps at (sub)mm wavelengths, and the authors had to rely
upon a miscellany of observations obtained with various telescopes. Now,
the situation has changed and we can take advantage of Galaxy-wide surveys
such as
the APEX Telescope Large Area Survey of the Galaxy (ATLASGAL; Schuller
et al.~\cite{schul09}), which covers almost all of the clumps studied by
FON02.

We recalculated the clump masses using the flux densities at $\nu=345$~GHz
from the ATLASGAL compact source catalogue (Urquhart et al.~\cite{urq14}; hereafter URQ14).
For the sake of consistency with FON02, we adopt their distances,
whereas we take the clump angular radius from URQ14
and $\To$ from Urquhart et al.~(\cite{urq18}; hereafter
URQ18). The latter is obtained from a modified black-body fit to the
SED and is hence a good approximation of
the temperature at the surface of the clump, because the SEDs of these
objects typically peak around $\sim$100~\mic\ where the emission is
optically thick. We also adopt $\kappa=1.85$~cm$^2$g$^{-1}$ and $\RR=100$
as in Schuller et al.~(\cite{schul09}), and assume $\ri=0.01$ based on the fact that
the density gradient with $p=-2.6$ appears to hold on a range of radii
spanning two orders of magnitude (see Fig.~10 of FON02).

The virial masses, $\Mv$, have been recalculated, using the line widths,
$\Delta V$, from FON02 and the new values of $\Ro$ and $\To$ from URQ14 and
URQ18. In our estimates, unlike FON02, we take into account the correction
to $\Mv$ due to the density and temperature profiles, as detailed in
Appendix~\ref{avir}.

\begin{figure}
\centering
\resizebox{8.5cm}{!}{\includegraphics[angle=0]{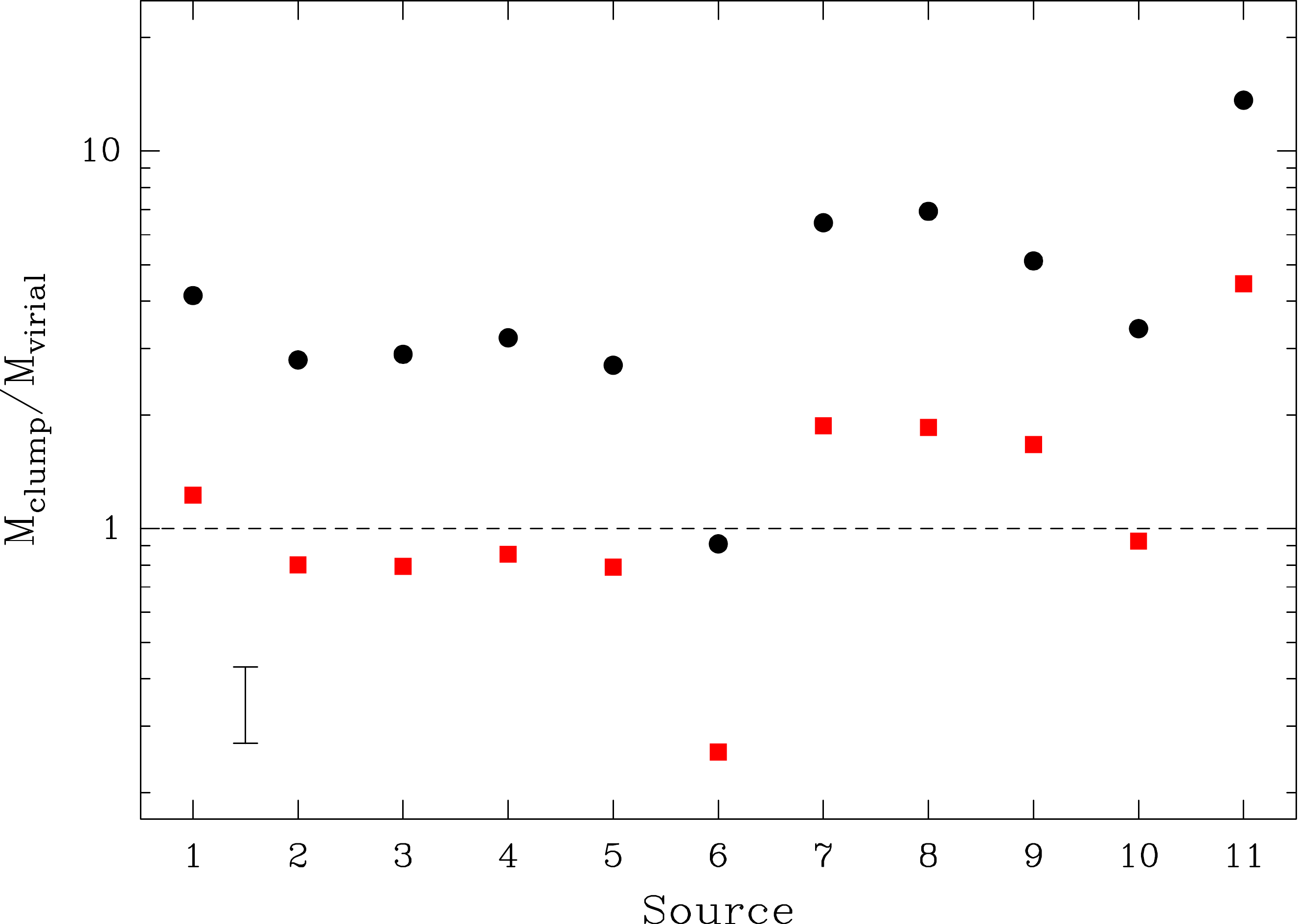}}
\caption{
Same as Fig.~6 of FON02, where the clump masses have been recomputed with
our numerical solution using the temperature, radii, and flux densities
from the ATLASGAL compact source catalogue, and the virial masses have been
corrected to take into account density and temperature gradients. The numbers
on the x-axis identify the clumps according to the numbering of Table~1
of FON02. Black circles correspond to constant density and temperature,
as assumed by FON02, whereas red squares are obtained adopting $q=-0.54$
and $p=-2.6$, consistent with the findings of FON02. The error bar in
the bottom left indicates the typical uncertainty on the mass ratio.
}
\label{fvir}
\end{figure}

Figure~\ref{fvir} is the same as Fig.~6 of FON02 and shows the ratio between
the clump mass and the corresponding virial mass for the different sources.
We have also evaluated a mean error on this ratio taking into account that
to a good approximation $M_{\rm clump}/\Mv\propto S_\nu/[\To\,(\Delta V)^2\,\tho]$
and assuming an uncertainty of 20\% for all variables. The plot confirms that basically
all clump masses are significantly greater than the corresponding virial
masses (black circles), if the clump mass is estimated with constant
temperature and density. However, when the temperature and density gradients
are taken into account with our numerical model, almost all clumps become
virialized (red squares). This result proves that the correction applied may
be crucial for stability issues.

\subsection{Masses of the ATLASGAL compact sources}

As a last example, we discuss how temperature and density gradients could
affect the estimates of the masses of the clumps identified in the ATLASGAL
compact source catalogue by URQ18. In particular, we calculate the ratio
between the mass computed with our method and that obtained by URQ18 from
Eq.~(\ref{eone}). For our estimates, $\tho$ and $S_\nu$ were taken from
Table~1 of URQ14, $\To$ and $d$ from Table~5
of URQ18, and we assume $\kappa(345~{\rm GHz})=1.85$~cm$^2$g$^{-1}$
and $\RR=100$ for consistency with URQ18. We also set $\ri=0.01$ for the
reason explained in Sect.~\ref{sfont}.

\begin{figure}
\centering
\resizebox{8.5cm}{!}{\includegraphics[angle=0]{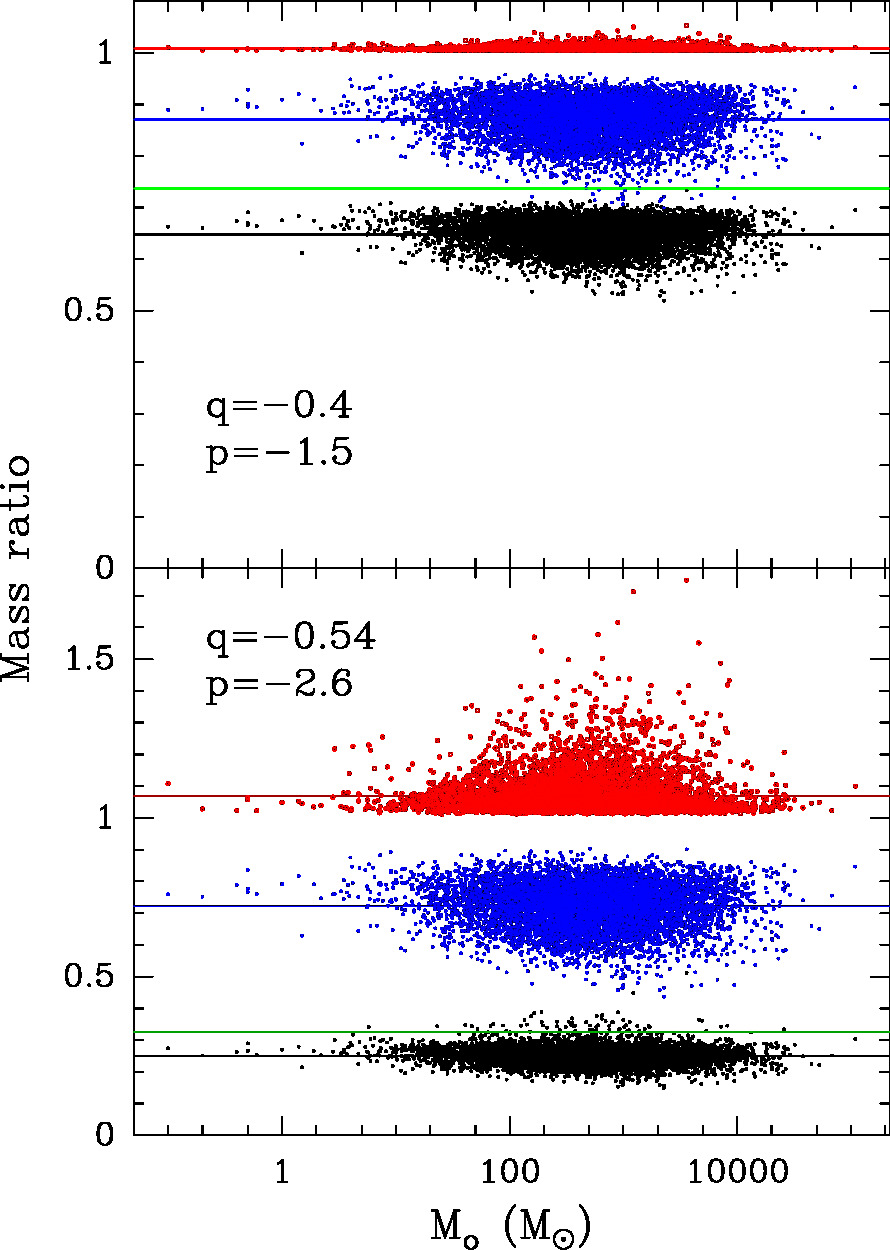}}
\caption{
The black dots indicate the ratio between the mass estimated with our
numerical model and that computed by URQ18 for the compact sources identified
in the ATLASGAL survey. The input parameters are taken from URQ18. For our
estimates, we assumed two template cases: $q=-0.4$, $p=-1.5$ (top panel)
and $q=-0.54$, $p=-2.6$ (bottom panel). The red and blue dots indicate,
respectively, the contribution of opacity and RJ approximation to the mass
ratio, with the horizontal lines denoting the corresponding mean values for
the black (0.64 top panel; 0.25 bottom panel), red (1.01 top panel; 1.07
bottom panel) and blue (0.87 top panel; 0.72 bottom panel) dots. The green
line is the factor (0.74 top panel; 0.33 bottom panel) taking into account
temperature and density gradients (see text for a detailed explanation).
}
\label{furq}
\end{figure}

\begin{table*}
\caption[]{
Approximate expressions of the flux density of a clump with density and
temperature gradients, in the RJ limit (for the definition of the symbols,
see Sects.~\ref{score} and~\ref{sdisk})
}
\label{teqs}
\centering
\begin{tabular}{l|cc}
\hline
\hline
opacity & $S_\nu$ spherical symmetry & $S_\nu$ cylindrical symmetry \\
\hline
$\tau\ll1$ & $\frac{\kappa M}{\RR d^2} \frac{2k\nu^2}{c^2}\To\, F(\ri;q,p+3)$
           & $ \frac{\kappa M}{\RR d^2} \frac{2k\nu^2}{c^2}\To\, F(\ri;q,p+2)$ \\
$\tau\gg1$ & $ \Omo\,\frac{2k\nu^2}{c^2}\To$
           & $ \frac{2k\nu^2}{c^2}\To \left( \Omo^{\rm e} \sin\psi + 2\,\Omo \cos\psi \int_{\ri}^1 r^{\,q+1} \d r \right)$ \\
\hline
\end{tabular}
\end{table*}

In Fig.~\ref{furq} we plot the ratio between our numerical mass estimate,
obtained as described in Sect.~\ref{snume}, and the mass computed by URQ18
(black dots), as a function of the latter ($\Moth$). The calculation is made
for the fiducial values of $q=-0.4$, $p=-1.5$ (top panel) and for $q=-0.54$,
$p=-2.6$ in the footsteps of Sect.~\ref{sfont} (bottom panel). The mean
ratio is, respectively, 0.64 and 0.25, which represent a non-negligible
correction for estimates of quantities such as e.g. the virial parameter.

It is instructive to examine the separate contributions of opacity, RJ approximation,
and temperature and density gradients to the correction factor. This can be done
by trivially re-writing the mass ratio as
\begin{equation}
 \frac{M}{\Moth} = \left(\frac{M}{\Mrj}\,\frac{\Morjth}{\Moth}\right) ~ \frac{\Mrj}{\Mrjth} ~ \frac{\Mrjth}{\Morjth}
 \label{emas}
\end{equation}
where the indices ``RJ'' and ``thin'' indicate, respectively, that the
mass is calculated in the RJ and in the optically thin approximation,
while the subscript ``o'' means that the calculation is done for constant
temperature and density (i.e. $T=\To$ and $\rho=\rhoo$).

In the right-hand side of Eq.~(\ref{emas}), the term in parentheses is
sensitive to the RJ approximation, the ratio $\Mrj/\Mrjth$ is related to
the opacity of the clump, and $\Mrjth/\Morjth=1/F$ is the correction
for the temperature and density gradients. These three quantities
are plotted in Fig.~\ref{furq} as red dots ($\Mrj/\Mrjth$), blue dots
($M\Morjth/(\Mrj\Moth)$), and a green line ($\Mrjth/\Morjth$).

We conclude that the most important correction is due to the gradients, although
in a non-negligible number of clumps opacity may play an important role, provided
the temperature and density gradients are sufficiently steep.

\section{Summary and conclusions}
\label{scon}

We have estimated the continuum emission from a dusty clump with temperature
and density gradients, assuming both spherical and cylindrical symmetry.
While our toy model assumes power-law profiles for the physical parameters,
it must be kept in mind that real clumps are more complex structures
where the temperature and density distributions are determined by heating
and cooling processes and must obey the laws of fluidodynamics. Also,
fragmentation and sub-clumpiness may affect the observed flux densities,
especially if coupled to large opacities. Finally, clumps are enshrouded
in more extended, lower density structures whose emission/absorption might
affect the measured flux from the clump. All these issues go
beyond the scope of our study, which is nonetheless useful to improve
on the usual simplified assumption of homogeneous, optically thin clumps.

We provide the reader with approximate analytical expressions (summarized
in Table~\ref{teqs}) to calculate the flux density as a function of the
clump mass and other relevant parameters and, conversely, derive the
mass from the measured flux in the optically thin and RJ limits. Also,
in Eqs.~(\ref{esol}) and~(\ref{erj}) we give an approximate solution
to the radiative transfer equation to calculate the brightness along
an arbitrary line of sight through the clump for any optical depth. Our
approach overcomes the problem represented by possibly steep density and
temperature gradients at small clump radii. The approximate solution is
then used to evaluate the flux density of the core numerically.

Comparison between the numerical and approximate analytical solutions
allows to inspect the limits due to the optically thin, Rayleigh-Jeans,
and constant-density/temperature approximations. We conclude that in most
cases the correction is about a factor 2--3, although in some extreme
cases characterised by unusually steep gradients and/or high frequencies,
the error introduced by the above approximations can be larger. In order
to illustrate all these effects, we have applied our method to three
practical examples taken from the literature, demonstrating that the
correction to the clump mass may significantly affect the estimate of
the clump stability.

\begin{acknowledgements}
It is a pleasure to thank Daniele Galli and Maite Beltr\'an for critically reading
the manuscript and useful suggestions.
\end{acknowledgements}

\begin{appendix}

\section{Function $F(\ri;q,a)$}
\label{afri}

The purpose of this appendix is to study the behaviour of $F$ defined by Eq.~(\ref{eff}) as
a function of $\ri$, in the non-trivial case $q\ne0$. In the following we consider three
possible cases depending on the value of $a$ and demonstrate that $F$ is always increasing
with $\ri$ if $q>0$, and decreasing if $q<0$.

\subsection{Case $a=0$}
\label{saz}

In this case $F(\ri)=(\ri^q-1)/\ln\ri^q$, which may be conveniently re-written as $F'(y)=(y-1)/\ln y$ with $y=\ri^q$.
The function is to be studied in the range $0<\ri\le1$ or $y>0$.

First of all we note that
\[
 \lim_{\ri\to0^+}{F(\ri)} =
 \left\{
 \begin{array}{lcl}
  \lim_{y\to0^+}{\frac{-1}{\ln y}} = 0 & \Leftrightarrow & q>0 \\
  \lim_{y\to+\infty} \frac{y}{\ln y} =  \lim_{t\to0^+} \frac{1}{-t\ln t} = +\infty & \Leftrightarrow & q<0
 \end{array}
 \right.
\]
and
\[
 \lim_{\ri\to1^-}{F(\ri)} = \lim_{y\to1}{F'(y)} = \lim_{t\to0} \frac{t}{\ln(1+t)} = 1
\]
where we have defined $t=y-1$.
Furthermore, the derivative of $F'(y)$ is equal to
\begin{equation}
 \frac{\d F'}{\d y} = \frac{y \ln y - y + 1}{y (\ln y)^2} = \frac{g(y)}{y (\ln y)^2}
\end{equation}
whose sign is determined by the sign of $g(y)=y \ln y - y + 1$.
Since $g(1)=0$ and
$\d g/\d y = \ln y>0 \Leftrightarrow y>1$, we conclude that $g$ has a minimum in $y=1$ and thus
$g\ge0$ for any $y>0$.
Consequently, $\d F'/\d y\ge0$ and $\d F/\d \ri = q \, \ri^q \, (\d F'/\d y)>0 \Leftrightarrow q>0$.

\subsection{Case $a\ne0$ and $a=-q$}

The result in this case is straightforward. Function
$F(\ri)=\ln\ri^a/(\ri^a-1)$ with $a=-q$ is the inverse of that studied
in Sect.~\ref{saz}, and is thus increasing with $\ri$ if and only if $a<0$, i.e. for $q>0$.

\subsection{Case $a\ne0$ and $a\ne-q$}

In this case it is convenient to re-write the function
\begin{equation}
 F(\ri)=\frac{a}{a+q} \frac{1-\ri^{a+q}}{1-\ri^a}
\end{equation}
assuming $y=\ri^a$ and $b=(a+q)/a$, which gives
\begin{equation}
 F'(y)=\frac{1}{b} \frac{1-y^b}{1-y}.
\end{equation}
For any value of $a\ne0$, one finds
\[
 \lim_{\ri\to1^-}{F(\ri)} = \lim_{y\to1}{F'(y)} = \lim_{t\to0} \frac{1}{b} \frac{1-(1+t)^b}{-t} = 1
\]
The calculation of the value of $F$ for $\ri=0$, depends on the sign of $a$. We obtain for $a>0$
\[
 \lim_{\ri\to0^+}{F(\ri)} = \lim_{y\to0^+}{F'(y)} =
 \left\{
 \begin{array}{lcl}
  \frac{1}{b} & \Leftrightarrow & b>0 \\
  +\infty & \Leftrightarrow & b<0
 \end{array}
 \right. \nonumber
\]
and for $a<0$
\[
 \lim_{\ri\to0^+}{F(\ri)} = \lim_{y\to+\infty}{F'(y)} =
 \left\{
 \begin{array}{lcl}
  \lim_{y\to+\infty} \frac{1}{b} \frac{y^b}{y} = +\infty & \Leftrightarrow & b>1 \\
  \lim_{y\to+\infty} \frac{1}{b} \frac{y^b}{y} = 0 & \Leftrightarrow & 0<b<1 \\
  \lim_{y\to+\infty} \frac{1}{b} \frac{1}{y} = 0 & \Leftrightarrow & b<0
 \end{array}
 \right. 
\]
We note that $b=0$ and $b=1$ are excluded because we are considering the
case for $a+q\ne0$ and $q\ne0$.

In conclusion,
\[
 \lim_{\ri\to0^+}{F(\ri)} =
 \left\{
 \begin{array}{lcl}
  0 & \Leftrightarrow & a<0,~q<0 \\
  \frac{a}{a+q} & \Leftrightarrow & a>0,~a+q>0 \\
  +\infty & \Leftrightarrow & a<0,~q>0 ~{\rm or}~ a>0,~a+q<0 \\
 \end{array}
 \right. \nonumber
\]

The derivative of $F$ is $\d F/\d\ri = a \ri^{\,a-1} \d F'/\d y$, where
\begin{equation}
 \frac{\d F'}{\d y} = \frac{1}{b} \frac{-b y^{b-1} (1-y) + 1-y^b}{(1-y)^2}
\end{equation}
so that the sign of $\d F/\d\ri$ depends on $ag(y)/b$, where we have defined $g(y)=-b y^{b-1} (1-y) + 1-y^b$.
We find that
$\d g/\d y=b(b-1)y^{b-2}(y-1)\ge0$ if $y\ge1$, for $b(b-1)>0$, and $y\le1$,
for $b(b-1)<0$. This means that $g$ has a minimum in $y=1$ if $b>1$
or $b<0$, a maximum if $0<b<1$. Consequently, for any $y$ it is $g\ge0$
in the former case and $g\le0$ in the latter, because in all cases $g(1)=0$.

Based on the above, one finds that $g/b>0 \Leftrightarrow b>1$,
so that $\d F/\d \ri \propto ag/b >0 \Leftrightarrow a>0,(a+q)/a>1~{\rm
or}~a<0,(a+q)a<1$.  Both conditions are equivalent to $q>0$. We conclude
that $F(\ri)$ is a growing function of $\ri$ if and only if $q>0$. Since
$F(1)=1$, this implies also that $F\ge1 \Leftrightarrow q<0$.

\section{Virial mass with density and temperature gradients}
\label{avir}

We want to derive the expression of the virial mass of a spherically
symmetric clump with temperature and density described by Eqs.~(\ref{etr}) and~(\ref{erh}).
The virial theorem can be expressed, e.g., as in Eqs.(8.4) and~(8.5) of Dyson \& Williams (\cite{dywi}),
namely
\begin{equation}
3\int P \d V = \int G \frac{M(R)}{R} \d M   \label{evir}
\end{equation}
where $P$ is the gas pressure, $V$ the volume, $M(R)$ the mass inside radius $R$, $G$
the gravitational constant, and we have assumed that the external pressure is null.
Using our notation (see Sect.~\ref{score}), $M(R)$ is obtained by integrating Eq.~(\ref{emass}) between $\Ri$ and $R$, i.e.
\begin{equation}
 M(R) = \int_{\Ri}^{R} \rho \RR \, 4\pi R'^2 \, \d R'  =  4\pi \Ro^3 \, \rhoo \RR \int_{\ri}^r r'^{p+2} \, \d r'
\end{equation}
which can be written as
\begin{equation}
 M(R) = M(\Ro) \frac{\int_{\ri}^r r'^{p+2} \, \d r'}{\int_{\ri}^1 r'^{p+2} \, \d r'}.
\end{equation}
The gas pressure is
\begin{equation}
 P(R) = \RR \rho(R) \left(\frac{kT(R)}{\mu} + \frac{\snt^2}{3}\right)
\end{equation}
where $\mu$ is the mean mass per particle and $\snt$ is the velocity dispersion due to microscopic non-thermal motions, which
we assume independent of $R$.
The virial mass, $\Mv$, is the value of $M(\Ro)$ that satisfies Eq.~(\ref{evir}), which takes the form
\begin{eqnarray}
 & & \hspace*{-5mm} \int_{\ri}^1 \left(3\RR\rhoo \frac{k\To}{\mu} r^{p+q} + \RR\rhoo \snt^2 r^p\right)\, \Ro^3 \, 4\pi r^2\, \d r   \nonumber \\
 & = &
 \int_{\ri}^1 G \Mv \frac{\int_{\ri}^r r'^{p+2} \, \d r'}{\int_{\ri}^1 r'^{p+2} \, \d r'} 4\pi\RR\rhoo \Ro^2 r^{p+1} \, \d r.
\end{eqnarray}
The solution is
\begin{eqnarray}
 \Mv & = & \frac{\snt^2 \Ro}{G} \int_{\ri}^1 r^{p+2} \, \d r \nonumber \\
 & & \times \, \frac{ \sr \int_{\ri}^1 r^{p+q+2} \, \d r + \int_{\ri}^1 r^{p+2} \, \d r }
 { \int_{\ri}^1 \left(\int_{\ri}^r r'^{p+2} \, \d r'\right) r^{p+1} \, \d r }
\end{eqnarray}
where we have defined $\so^2=3k\To/\mu$ and $\eta=\so^2/\snt^2$. Depending on the values of $p$ and $q$
the solution takes the following forms:
\begin{eqnarray}
 \Mv & = & \Mnt \nonumber \\
 & & \hspace*{-14mm} \times
 \left\{
 \begin{array}{lcl}
  \left(\sr+1\right)\,\frac{\ri\,\ln^2\ri}{1+\ri\,\ln\ri-\ri} & \Leftrightarrow & p=-3,~q=0 \\
  \ri\,\ln\ri \frac{\ln\ri-\sr\frac{1-\ri^q}{q}}{1+\ri\,\ln\ri-\ri} & \Leftrightarrow & p=-3,~q\ne0 \\
  \left(\frac{1-\ri^{p+3}}{p+3}-\sr\ln\ri\right) \frac{1-\ri^{p+3}}{E(\ri,p)} & \Leftrightarrow & p\ne-3,~q=-p-3 \\
  \left(\sr\frac{1-\ri^{p+q+3}}{p+q+3}+\frac{1-\ri^{p+3}}{p+3}\right) \frac{1-\ri^{p+3}}{E(\ri,p)} & \Leftrightarrow & p\ne-3,~q\ne-p-3 \\
 \end{array}
 \right.
\end{eqnarray}
where $\Mnt=\snt^2\Ro/G$ and
\begin{eqnarray}
 E(\ri,p) & = & \int_{\ri}^1\left(r^{p+3}-\ri^{p+3}\right) r^{p+1} \,\d r \nonumber \\
 & = & \left\{
 \begin{array}{lcl}
 -\ln\ri + 2\frac{\ri^\frac{1}{2}-1}{\ri} & \Leftrightarrow & p=-\frac{5}{2} \\
 1-\ri+\ri\ln\ri & \Leftrightarrow & p=-2 \\
 \frac{1-\ri^{2p+5}}{2p+5} + \frac{\ri^{2p+5}-\ri^{p+3}}{p+2} & \Leftrightarrow & p\ne-\frac{5}{2},~p\ne-2 \\
 \end{array}
 \right.
\end{eqnarray}

It is possible to demonstrate that if $p\le-5/2$ or $q\le-p-3$, for
$\ri\to0^+$ no equilibrium configuration can be attained, because either
the gravitational energy overwhelms the internal energy of the clump
($\Mv\to0$) or the opposite happens ($\Mv\to+\infty$). Vice versa, for
$p>-5/2$ and $q>-p-3$ one finds that
\[
\lim_{\ri\to0^+}{\Mv} = \Mnt \, (2p+5) \left(\frac{1}{p+3}+\frac{\sr}{p+q+3}\right)
\]
which for $\sr=0$ (i.e. negligible thermal contribution to the internal
energy) turns into Eq.~(1) of MacLaren et
al.~(\cite{maclar})\footnote{These authors erroneously state that their
equation holds for any $p>-3$, instead of $p>-5/2$.}.

The mass $\Mnt$ can be conveniently expressed in useful units as
\begin{equation}
 \Mnt = \frac{3\,(\Delta V)^2}{8\,\ln2} \frac{\Ro}{G} = 125.8~M_\odot ~ [\Delta V(\kms)]^2 \, \Ro({\rm pc})
\end{equation}
where the factor 3 takes into account that the observed line full width
at half maximum, $\Delta V$, is a measurement of the velocity dispersion
along the l.o.s., i.e. in one dimension.

The relevant parameters for the case discussed in Sect.~\ref{sfont} are
$p=-2.6$, $q=-0.54$, and $\ri=0.01$, which imply $\Mv\simeq\Mnt (1.496\,\sr+0.487)$,
with $\sr\simeq0.01647\,\To(K)/[\Delta V(\kms)]^2$. Here we have assumed
$\mu=2.8\,m_{\rm H}$, with $m_{\rm H}$ mass of the hydrogen atom.

\end{appendix}


\begin{thebibliography}{}

\bibitem[2018]{bel18}
 Beltr\'an, M. T., Cesaroni, R., Rivilla, R. et al. 2018, A\&A, 615, A141 (BEL18)
\bibitem[2010]{cesa10}
 Cesaroni, R., Hofner, P., Araya, E., \& Kurtz, S. 2010, A\&A, 509, A50
\bibitem[2019]{cesa19}
 Cesaroni, R., Beltr\'an, M. T., Moscadelli, L., S\'anchez-Monge, \'A, \& Neri, R. 2019, A\&A, 624, A100
\bibitem[1980]{dywi}
 Dyson, J.E. \& Williams, D.A. 1980, The Physics of the Interstellar Medium, Manchester University Press, Manchester
\bibitem[2002]{fonta02}
 Fontani, F., Cesaroni, R., Caselli, P., \& Olmi, L. 2002, A\&A, 389, 603 (FON02)
\bibitem[1983]{hild83}
 Hildebrand, R. H. 1983, QJRAS, 24, 167
\bibitem[1988]{maclar}
 MacLaren, I., Richardson, K. R., \& Wolfendale, W. 1988, ApJ, 333, 821
\bibitem[1994]{osshen}
 Ossenkopf, V. \& Henning, Th. 1991, A\&A, 291, 943
\bibitem[2009]{schul09}
 Schuller, A., Menten, K. M., Contreras, Y., et al. 2009, A\&A, 504, 415
\bibitem[2014]{urq14}
 Urquhart, J. S., Csengeri, T., Wyrowski, F., et al. 2014, A\&A, 568, A41 (URQ14)
\bibitem[2018]{urq18}
 Urquhart, J. S., K\"onig, C., Giannetti, A., et al. 2018, MNRAS, 473, 1059 (URQ18)

\end{thebibliography}
\end{document}